\documentclass[12pt]{article}
\usepackage[body={17.5cm, 22cm},right=2cm]{geometry}
\usepackage{amssymb,graphicx}
\usepackage{amsmath}
\usepackage{epsfig}
\usepackage{hyperref} 

\def\e{{\rm e}}

\def\d{\partial}
\def\l{\left(}
\def\r{\right)}

\newcommand{\be}{\begin{equation}}
\newcommand{\ee}{\end{equation}}
\newcommand{\bea}{\begin{eqnarray}}
\newcommand{\eea}{\end{eqnarray}}
\newcommand{\bg}{\begin{gather}}
\newcommand{\eg}{\end{gather}}
\newcommand{\bseq}{\begin{subequations}}
\newcommand{\eseq}{\end{subequations}}

\newcommand{\Tr}{{\rm Tr}}

\begin{document}
\begin{center}
{\Large\bf Black Hole Portal into Hidden Valleys}\\
\vspace{0.5cm}
{ \large
 Sergei~Dubovsky, Victor Gorbenko}\\
\vspace{.2cm}
{\small  \textit{Center for Cosmology and Particle Physics, Department of Physics, \\New York University, New York, NY, 10003, USA }}

\end{center}
\begin{center}
\begin{abstract}
Superradiant instability turns rotating astrophysical black holes into unique probes of light axions. We consider what happens when a light axion is coupled to a strongly coupled hidden gauge sector. In this case superradiance results in an adiabatic increase of a hidden sector CP-violating $\theta$-parameter in a near horizon region. This may trigger a first order phase transition in the gauge sector. As a result a significant fraction of a black hole mass is released as a cloud of hidden mesons and can be later converted into electromagnetic radiation.
This results in a violent electromagnetic burst. The characteristic frequency of such bursts may range from $\sim 100$ eV to $\sim 100$ MeV.

\end{abstract}
\end{center}
%
\section{Introduction}
String theory landscape points towards ultimate unification of particle physics and geography. 
If the landscape is real it is very likely that in the short and midterm future  particle physicists will be busy by mapping  the compactification manifold where we happened to live. In the most optimistic scenario---if extra dimensions are large---this can be achieved directly at the LHC and future colliders.
However,  even if extra dimensions are small they are still likely to be populated by rich structures accessible for collider, laboratory, astrophysical and cosmological probes at relatively low
energies. These structures escaped detection so far as a consequence  of their distant geographic location in extra dimensions.

The structures which are natural to expect in the landscape on theoretical grounds include hidden gauge sectors (``hidden valleys", \cite{Strassler:2006im}) and a plenitude of axion-like particles, dark photons and photini
(the ``axiverse", \cite{Arvanitaki:2009fg}). Confinement scales of strongly coupled gauge sectors and axion masses are exponentially sensitive to the parameters of the compactification manifold  and  may  take arbitrarily low values. For
 axions there is also an observational indication that they may be light. Indeed, not only the QCD contribution to the mass of the QCD axion is very small ($\sim 10^{-10}$~eV for the 
GUT scale axion decay constant), but also additional string corrections to the axion potential, expected to be there on general grounds, have to be  suppressed  at least by extra ten orders of magnitude. Consequently, one may expect the presence of even lighter axions with masses dominated by these corrections.

A lot of efforts were concentrated recently on exploring the possibilities to discover these structures in the near future experiments and observations (see, e.g.,  \cite{Strassler:2006im}--\cite{Cheung:2010mc}). The focus of the current work will be the intriguing possibility
to probe light axion particles with masses in the range $10^{-10}\div10^{-20}$~eV with ongoing observations of astrophysical black holes \cite{Arvanitaki:2009fg,Arvanitaki:2010sy}. This possibility is related to the famous Penrose process allowing extraction of kinetic energy from rotating black holes \cite{Penrose:1969pc}--\cite{Starobinskii}. If a boson with a Compton wavelength of order the size of a black hole is present, the Penrose process results in a superradiant instability \cite{Damour:1976kh}--\cite{Detweiler:1980uk}. A black hole releases its rotational energy by creating a cloud of a rotating Bose--Einstein condensate in the near horizon region. 
Compton wavelengths of axions in the above mass range match sizes of astrophysical black holes and
Ref.~\cite{Arvanitaki:2010sy} studied opportunities provided by the superradiant instability for the discovery of these particless. The focus of Ref.~\cite{Arvanitaki:2010sy} was mainly on model-independent  purely gravitational signatures of  superradiance. Perhaps one of the most interesting conclusions is that Advanced LIGO may be able to discover the QCD axion if the decay constant is high.

Instead, here we will consider model-dependent signatures  arising when an axion causing the instability is coupled to a QCD-like hidden gauge sector (a hidden valley). The presence of these signatures depends on some of the details of the matter spectrum in the hidden valley. The payoff that we get for introducing a certain amount of model-dependence is that the resulting signals are electromagnetic rather than purely gravitational.

Concretely, we are exploiting the following effect. From the point of view of a hidden gauge sector a growing cloud of axions surrounding a rotating black hole acts as a QCD laboratory with a slowly oscillating CP-violating $\theta$-parameter. The amplitude of oscillations grows and reaches order one at the late stages of the instability. As we discuss in section~\ref{phases}, it is a very common feature of gauge theory dynamics that under an adiabatic increase of $\theta$ the vacuum which is stable at $\theta=0$  first  turns into an overheated metastable minimum and later on either becomes perturbatively unstable or even totally disappears as an extremum of the potential. 

Unfortunately, this possibility is not realized in the real-world QCD. However, it may be exhibited in one of hidden valleys. If this happens, the superradiant instability will trigger an avalanche phase transition in the hidden valley and a cloud of strongly interacting mesons will get produced. The total energy of the cloud depends on the axion decay constant, and for high scale axions may constitute $10^{-4}\div 10^{-3}$ of a black hole mass. Unlike axions, whose couplings to the rest of the matter are strongly suppressed by the decay constant, hidden sector mesons may have substantial interactions with the Standard Model fields. The net result is that a significant fraction of a black hole mass can be converted into electromagnetic energy on a characteristic time-scale set by a black hole size. This would give rise to violent transient astronomical sources---electromagnetic echoes of hidden valley avalanches---that can be looked for in a wide range of frequencies. 

The rest of the paper is organized as follows. In section~\ref{phases} we review how a vacuum structure with a non-trivial dependence on a $\theta$-parameter
may arise in a QCD-like theory. As a fully calculable example we use a QCD-like theory with $N$ light quark flavors, where the relevant dynamics can be analyzed with the help of a low energy chiral Lagrangian. In section~\ref{avalanches} we discuss in details how superradiant instability may trigger an avalanche phase transition in a hidden valley. 
In section~\ref{EM} we estimate characteristic energies, frequencies and time-scales of the corresponding electromagnetic signal as a function of an axion mass. In the concluding section~\ref{conclusions} we present our conclusions and outline directions for the future work.
\section{Phase Structure of Hidden Valleys}
\label{phases}
To illustrate how an adiabatic change of a $\theta$-parameter can trigger a phase transition in a strongly coupled gauge sector, let us consider $SU(N_c)$ QCD-like gauge theory with $N$ light quark flavors (this section is mostly a summary of well-known facts, an incomplete list of original references include  \cite{Witten:1980sp,Creutz:1995wf,Smilga:1998dh}).
In this case all the relevant dynamics can be analyzed in a weak coupling regime using the low energy chiral Lagrangian.
Indeed, from the chiral perturbation theory we get the following effective 
Lagrangian
\begin{equation}\begin{split} 
  	{\cal L} ={F_\pi^2\over 4}\Tr\,\d U^\dagger\d U +
  	             \frac{ \Lambda^{3}}{2}
  	                 Tr \left( M \e^{-i\theta /{N}} U + M \e^{i\theta /{N}} U^{\dagger} \right) 
  	                ,
  \label{V(U)}
\end{split}\end{equation}
where 
$ U\in SU(N)$ are pions,
$ M$ 
is the quarks mass matrix and the pion decay constant $F_\pi$ is of the same order as the QCD scale $\Lambda$. With appropriate field redefinitions one can always make $M$ real and diagonal, $ M = diag \{ m_1,  m_2 \ldots m_N \} $, with 
\be
\label{massorder}
 m_1 \geq m_2 \geq \ldots m_N >0\;.
 \ee 
 All masses are assumed to be much smaller than the dynamical  QCD scale $\Lambda$~\footnote{Strictly speaking, in  the large $N_c$ regime, one needs $m_i\ll \Lambda/N_c$ for the description (\ref{V(U)}) to hold, otherwise one should also include the $\eta'$ meson in the low energy theory.}.

Let us  study  the  extrema of the potential (\ref{V(U)});  our main interest is to follow the dependence on the CP-violating phase $\theta$. 
As we show in the Appendix, one can always search for  extrema in the diagonal form, $U=\e^{i\theta/N}diag \{ e^{i \phi_1},e^{i \phi_2}, \ldots e^{\phi_{N}} \}$, with
\be
\label{sum0}
\sum_i^N\phi_i=-\theta+2\pi k\;,
\ee
where $\phi_i\in [-\pi,\pi)$ and $k$ is an integer.
With this diagonal Ansatz the pion potential becomes
\begin{equation}\begin{split} 
  	V(\phi_i) = - \Lambda^3  \sum_i^N m_i cos\phi_i
  \label{V(a)}\;.
\end{split}\end{equation}
%
The extremality conditions for the potential (\ref{V(a)}) are simply
\begin{equation}\begin{split} 
  	  m_i sin{ \phi_i} =m_j sin \phi_j
  \label{extrEq},
\end{split}\end{equation}
for any pair $i,j$, plus the constraint (\ref{sum0}). For each value of  $\theta$ this set of equations has various
solutions which correspond to different type of extrema --- minima,
maxima and saddle points. 

\begin{figure}[t!] 
 \begin{center}
 \includegraphics[width=3.3in]{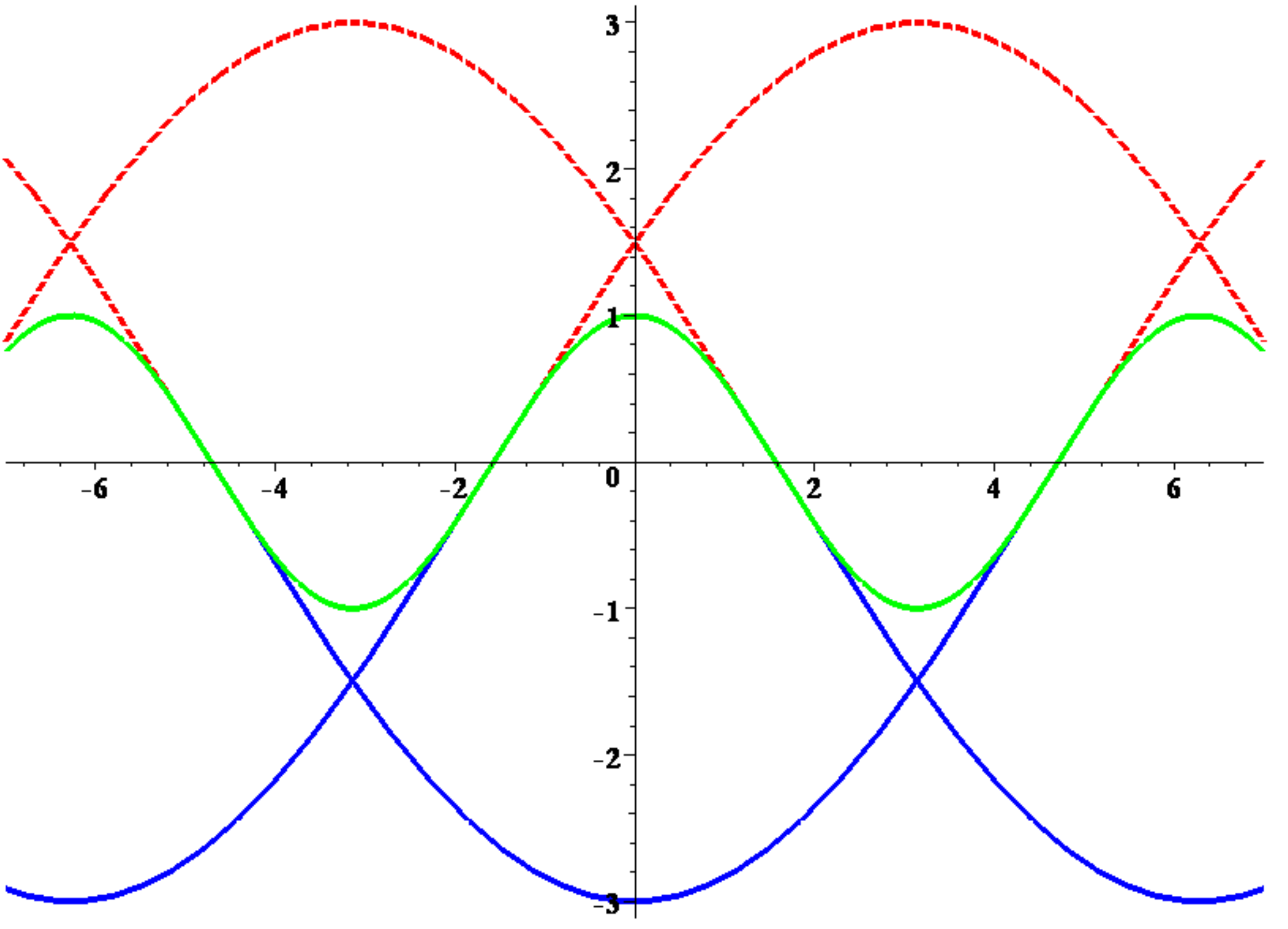}
 \includegraphics[width=3.3in]{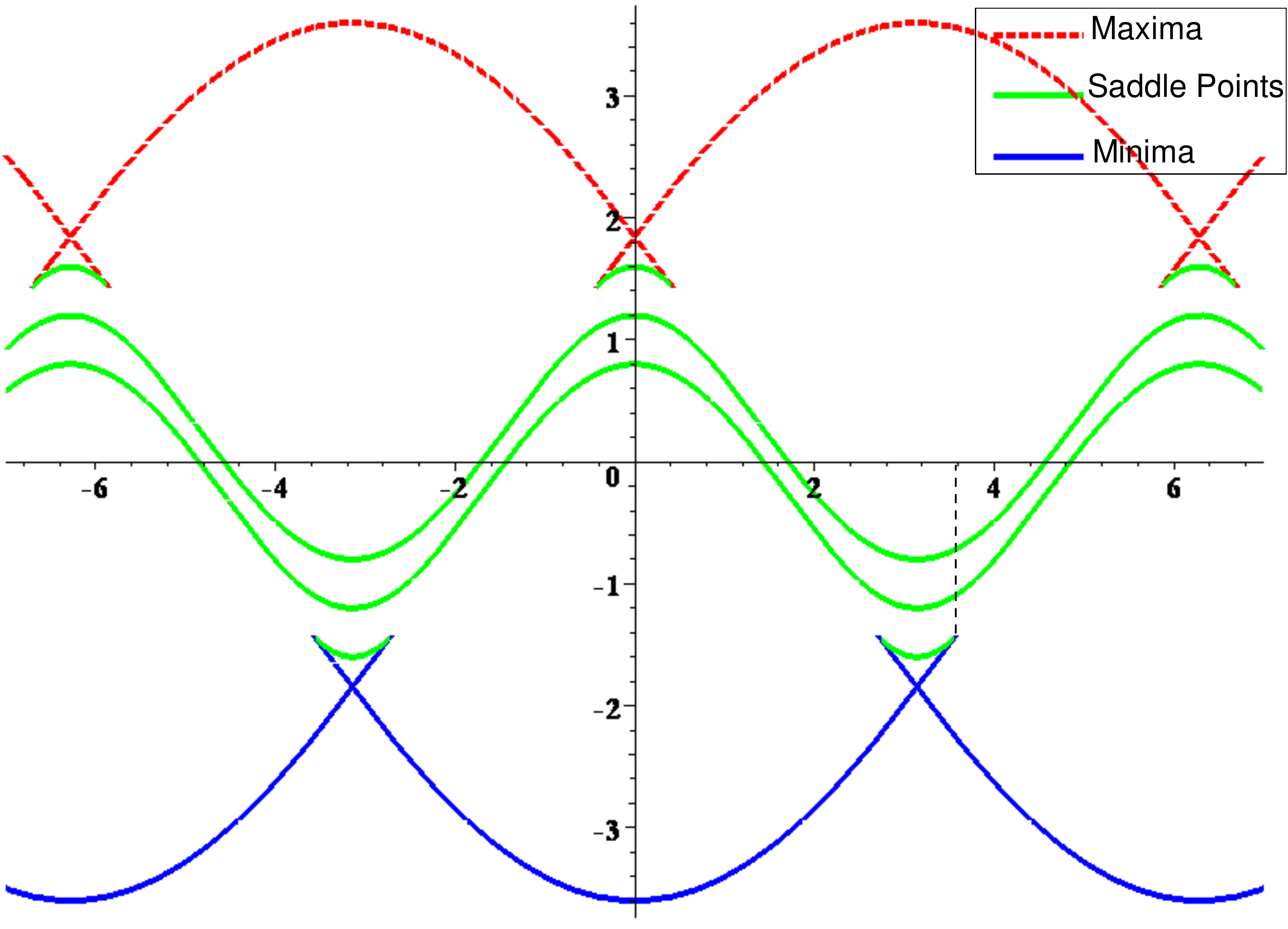}
 \put(-360,-10){a)}
  \put(-115,-10){b)}
 \put(-10,92){$\theta$}
 \put(-64,92){$\theta_c$} 
 \put(-357,170){$V$}
 \put(-252,92){$\theta$}
 \put(-294,92){$\theta_c$} 
 \put(-115,170){$V$}
 \put(-287.5,86){\bf\Large .}
  \put(-64.5,53){\bf\Large .}
 \caption{Extrema of the axion potential at $N=3$ for equal quark masses (left), and for mass ratios 1:1.2:1.4 (right).}
 \label{fig:vacua}
 \end{center}
\end{figure}
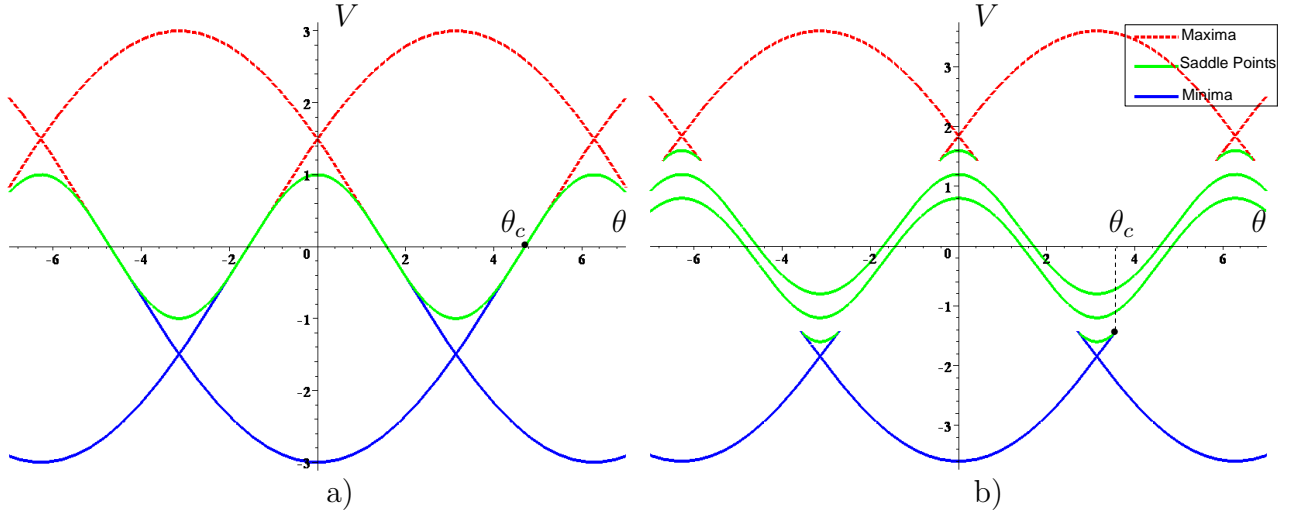

To determine conditions for an extremum to be a local minimum let us 
consider the Hessian matrix ${\cal M}_{\alpha\beta}$ using $\phi_1,\dots,\phi_{N-1}$ as independent variables, so that indices $\alpha,\beta$ run from $1$ to $(N-1)$,
\begin{equation}\begin{split} 
{\cal M}_{\alpha\beta}=\delta_{\alpha\beta}m_\alpha\cos\phi_\alpha+m_N\cos\phi_N
  \label{hessian},
\end{split}\end{equation}
where no summation over $\alpha$ is assumed. Then it follows from Eqs.~(\ref{massorder}), (\ref{extrEq}) that a necessary condition for a local minimum is 
$\cos\phi_\alpha>0, \ \alpha=1, \ldots, N-1$. Moreover, if also $\cos\phi_N>0$ then the extremum is necessary a minimum, while for negative  $\cos\phi_N$ it can be both stable or unstable.
 For equal quark masses
$m_i\equiv m$, this immediately implies that the minima correspond to the solutions of (\ref{sum0}), (\ref{extrEq}) where all $\phi_i$ are equal to each other,
\be
\label{sols}
\phi_i={-\theta+2\pi k\over N}\;,
\ee
and the parameter $k$ should be in the range
\be
\label{krange}
-{N\over 4}+{\theta\over 2\pi}<k<{N\over 4}+{\theta\over 2\pi}
\ee
for an extremum to be a minimum. The value of the axion potential in these minima is 
\be
\label{vacen}
V_k=-N\Lambda^3m\cos\l{\theta-2\pi k\over N}\r\;.
\ee
At $\theta=0$ there are $(1+2[N/4])$ minima, with $k=0$ corresponding to the global one. Let us follow now what
happens with the energy if one starts at the global minimum at $\theta=0$ and adiabatically increases the value of $\theta$. As we will see in the next section this is exactly what happens in the vicinity of a rotating astrophysical black hole if $\theta$ is promoted to a dynamical axion field with an appropriate mass.
It follows from (\ref{vacen}) that  the energy $V_0$ of the $k=0$ minimum increases with $\theta$. At $\theta=\pi$ this vacuum ceases to be a global minimum. Finally, there is a critical value $\theta_c$, such that for $\theta>\theta_c$ the $V_0$ branch of extrema does not correspond  even to a local minimum of energy any longer. In the case of equal quark masses
\[
\theta_c={N\pi\over 2}
\]
and $V_0$ is a local maximum for $\theta>\theta_c$. In fact, for equal masses it is straightforward to classify all the extrema of the pion potential. In Fig.~\ref{fig:vacua}a we reproduced the plot from \cite{Smilga:1998dh}, where all of the extrema are shown for $N=3$. We see that indeed the branch of extrema corresponding to the global minimum at $\theta=0$ (for $N=3$ this is actually the only minimum at $\theta=0$) becomes metastable at $\theta=\pi$, and turns into a local maximum after hitting a saddle point at $\theta=3\pi/2$.

A number and properties of extrema change continuously under small variation of the parameters  of 
the potential at a generic value of $\theta$. Hence also at non-degenerate masses adiabatic variations of $\theta$ may first  turn a global minimum into a metastable one,  and as the deviation of $\theta$ from zero increases, may  result in a disappearance of a minimum, as soon as the mass splittings are not too large.  For instance, for three quark flavors this behavior persists 
as soon as $(m_2-m_3)m_1 \leq m_2 m_3$. Unfortunately, this condition is not satisfied in the real world QCD. 

To illustrate how the structure of extrema changes under variations of masses 
we presented in Fig.~\ref{fig:vacua}b different extrema as a function of $\theta$ for mass ratios 1:1.2:1.4.
There is still a range of $\theta$ where a branch corresponding to the global vacuum at $\theta=0$ is a local minimum. 

 There is one interesting difference of this plot from the one in  Fig.~\ref{fig:vacua}a. $\theta$ is always a periodic parameter with a period equal  to 2$\pi$. The way this periodicity is realized at equal masses is that as $\theta$ varies adiabatically from 0 to 2$\pi$ the set of extrema gets reshuffled. This effect is often referred to as axion monodromy. Even at equal masses this name is a bit misleading---as we see on the way from 0 to $2\pi$ extrema change their nature and turn from minima into maxima and vice versa. Fig.~\ref{fig:vacua}a demonstrates that for non-degenerate masses
 something even more dramatic happens---there are branches of extrema that are not  defined 
 over all range of $\theta$ from 0 to $2\pi$. Of course, the physics is still periodic under $2\pi$ shifts of $\theta$---the sets of extrema at $\theta$ and $\theta+2\pi$ are identical.
 
 Note that the two minima have equal energies at $\theta=\pi$ both for different and degenerate masses.
This is not a coincidence, but rather a consequence of the $2\pi$-periodicity in $\theta$ and of  $\theta\to -\theta$ reflection symmetry.

Let us estimate now the life-time of metastable minima in the pion potential. As usual the probability of tunneling is determined by the classical bounce action, $\Gamma\propto \e^{-S}$ \cite{Kobzarev:1974cp,Coleman:1977py}. 
It is clear that the tunneling probability becomes large for $\theta$ close to $\theta_c$, where the height of the barrier goes to zero. Indeed the relevant part of the pion Lagrangian at  $\theta\sim\theta_c$
can be written in the form
\begin{equation}\begin{split} 
{\cal L}\simeq {F_\pi^2\over 2}\l\d\phi\r^2-m \Lambda^3 \left(   M^2(\theta) \phi^2 - 
                        \lambda \phi^3 +\dots \right),
  \label{cubicPot}
\end{split}\end{equation}
where $M^2(\theta_c)=0$, $\phi$ is a linear combination
of pions $\phi_i$ that becomes massless at $\theta=\theta_c$,  and $\lambda$
is a constant. Dots in (\ref{cubicPot}) stand for  corrections which are higher order in $\phi$ and/or in $(\theta-\theta_c)$.
These terms can be neglected at $\theta\sim\theta_c$ when the bounce solution is localized near $\phi=0$\footnote{The cubic coupling $\lambda$ can turn zero at some special values of quark masses. In this case one has to include the quartic self-interaction to calculate the bounce action. This will change some of our formulas but not the conclusions.}. In this limit the tunneling rate becomes \cite{Linde:1981zj}
%
%
\begin{equation}\begin{split} 
\Gamma \sim \exp\l{- {91\over 2}\frac{M^2(\theta)}{\lambda^2}\frac{F_{\pi}^4}{m \Lambda^3} }\r\;,
  \label{G}
\end{split}\end{equation}
{\it i.e.}, the tunneling is indeed unsuppressed for $\theta\sim\theta_c$.
%
It will be important for what follows that this tunneling regime corresponds to a thick wall limit, when a significant energy $\sim m\Lambda^3$ is released in pions as a result of a tunneling.
For this regime to be physically relevant we need to check that the tunneling is slow in the opposite  (thin wall) regime.
The tunneling probability in the thin wall limit is given by \cite{Coleman:1977py}
%
%
%
\begin{equation}\begin{split} 
\Gamma\sim\exp\l{-\frac{27}{2} \pi^2 \frac{\sigma^4}{\Delta E^3}}\r\;,
  \label{Coleman}
\end{split}\end{equation}
where $\Delta E$ is the energy difference between the vacua 
and $\sigma$ is the domain wall tension.
To estimate $\sigma$ we  consider the equal mass case when
the transition proceeds  between vacua with $U={\mathbf 1}$ and $U=e^{i 2 \pi / N} {\mathbf 1}$.
The domain wall tension $\sigma$ is then equal to the Euclidean action functional per unit surface area
\begin{equation}\begin{split} 
\int_{-\infty}^{\infty} dx\;\l \frac{F_{\pi}^2}{4} \Tr\, \d_xU^{\dagger}\d_xU +V\r
  \label{E}
\end{split}\end{equation}
calculated on the optimal solution, which depends on a single  variable $x$.
%
%
It was calculated for three quark flavors in \cite{Smilga:1998dh}. We will estimate the tension for a general $N$
by assuming the same Ansatz 
$U(x)=diag \{ \e^{i \gamma(x)},\e^{i \gamma(x)} \ldots \e^{-i(N-1) \gamma(x)} \}$.
The result is
\begin{equation}\begin{split} 
\sigma= F_{\pi} \sqrt{m \Lambda^3} I_N
  \label{sigma}
\end{split}\end{equation}
where $I_N$ is a numerical coefficient  of order 1 (for $N=3$  it is equal to $6(1-\pi/(3\sqrt{3})$).
This results in the following estimate for the tunneling rate,
\begin{equation}\begin{split} 
\Gamma \sim  \exp\l{- \frac{27 \pi^2 I_N^4}{16} \frac{F_\pi^4}
                     {m \Lambda^3 \sin^3 \left ( \frac{\pi}{N} \right) (\theta-\pi)^3 }}\r\;,
  \label{p}
\end{split}\end{equation}
We see that the thin wall rate is naturally suppressed both by a large numerical factor in the exponent, and also
because the ratio $F_\pi/m$ is large in a chiral limit.
We conclude  
that the transition should be close to the thick wall regime in order to proceed efficiently, giving rise to a significant latent heat  $\sim m\Lambda^3$, which is released in the form of pions.

To finish this section note that a non-trivial vacuum structure as a function of $\theta$ parameter of the type discussed here appears to be a generic property of strongly coupled gauge theories, not limited to a specific calculable example that we chose. In particular, it is likely to be a feature of the pure gluodynamics as well \cite{Witten:1998uka,Shifman:1998if}.
\section{Hidden Valley Avalanches}
\label{avalanches}
Let us describe now how axionic superradiant instability of astrophysical black holes may trigger rapid  instabilities  (avalanches) in hidden valleys. Eventually, our main focus will be on possible electromagnetic echoes of these avalanches, but in this section we describe the relevant dynamics entirely within a hidden sector.
Let us start with a quick review of the basics of superradiant instability following Ref.~\cite{Arvanitaki:2010sy}. 

Light axion fields render rotating astrophysical black holes unstable. The instability rate depends on the ratio $\alpha$ of the black hole size $r_g$ and the axion Compton wavelength, $\alpha=\mu_a r_g$. For a maximally spinning black hole the rate is maximum, $\Gamma\sim 10^{-7}r_g^{-1}$, at $\alpha\approx 1/2$, and rapidly decreases when $\alpha$ deviates from this optimal  value. In practice  the size of astrophysical black  holes is tiny compared to the age of the Universe, so that $\alpha$ can be significantly different from its optimal value and still the instability has enough time to develop. Every axion may affect black holes within two-three orders of magnitude in mass (depending on $\mu_a$), as illustrated in Fig.~2 of  Ref.~\cite{Arvanitaki:2010sy}. 

 It is useful to think about black hole/axion system as a huge gravitational atom. One crucial difference with a hydrogen atom is that axions 
are bosons so that  energy levels may have large occupation numbers (of order $10^{70}$ in the present context).   Another important difference is that an atomic nucleus is replaced by a black hole, as a result the boundary condition at the origin gets modified, and the energy levels acquire imaginary parts. For a Schwarzschild black hole all imaginary parts are negative, implying that a black hole absorbs particles from all levels.
For a rotating black hole some of the levels acquire positive imaginary parts, signaling the presence of a superradiant instability. The occupation numbers of these superradiant levels grow exponentially by extracting rotational energy from the ergosphere. This is a variation of the famous Penrose process.

To describe the development of the instability let us start by considering a black hole which rotates rapidly at the initial moment of time. This may  be a realistic situation if, e.g.,  a  black hole is just created as a result of a supernova explosion with a large natal spin. 

From that moment on the black hole starts rapidly loosing its spin by populating the fastest superradiant level. The process continues until  non-linearities in the axion potential become important. For an axion potential of the form
shown in Eq.~(\ref{vacen}) (with $k=0$) this happens at $\theta/N\sim 1$. In the absence of a hidden gauge sector the estimates of  Ref.~\cite{Arvanitaki:2010sy} suggest that as an outcome 
of  non-linear dynamics the axion cloud collapses---a similar process was observed for attractive Bose--Einstein condensates in a lab, and is called ``Bosenova" \cite{Bosenova}. After the collapse, presumably, an order one fraction of axions gets absorbed back into the black hole, and an order one fraction escapes. After the environment cleans up the whole process repeats. This happens until the black hole spin drops down to the critical value, where the width of the fastest superradiant level 
changes its sign. 

At this point one may expect that the next-to-fastest superradiant level becomes dynamically important and the whole story repeats with a slower time-scale set by its population rate.  However, the presence of the axion cloud populating the first level induces mixing between other levels that shuts down the instability for these levels. As a result the process repeats for the second level only after a cosmologically long time required for the axion cloud to dissipate, so that the level mixing becomes sufficiently small. In the absence of external perturbations the dissipation happens due to axion annihilations into gravitons. Alternatively, the cloud can be destabilized as a consequence
of a merger event, or if the black hole mass and spin changes by order one as a result of a long period of accretion.

In all these cases a time interval between episodes of superradiance instability is cosmologically long, so that every black hole may experience several such episodes through the history of the Universe. Interestingly, 
these episodes may be happening for isolated black holes that do not correspond to any visible astronomical sources before the superradiance starts.
The focus of  Ref.~\cite{Arvanitaki:2010sy}, was mainly on possible gravitational wave signals during the episodes of superradiance for a generic axiverse axions and also on the possible non-gravitational signals for the QCD axion.
Instead, here we consider what happens if an axion causing the superradiance is coupled to the QCD-like hidden sector with the phase structure as described in section~\ref{phases}.
For concreteness, we consider the case of equal quark masses. It is straightforward to extend the whole discussion to 
the case of different but close enough masses, such that the non-trivial phase structure persists.

In the field theoretic language the growth of the axion cloud implies that the value of the axion field gets excited from the minimum of its potential in the vicinity of a black hole. It  experiences oscillations with a period set by the axion mass, and with an exponentially growing amplitude. The spatial profile of the field follows the shape of the hydrogen wave function with a typical variation scale set by the Bohr radius, $R_c\sim {l^2/(\alpha \mu_a)}$, where $l$ is the angular momentum of the corresponding superradiant level. 

Note that  the scales for both time and space variations of the axion field are extremely slow compared to the dynamical scale $\Lambda$ of the corresponding strongly coupled 
gauge sector. Consequently, from the hidden valley viewpoint the interior of the axion cloud can be thought of as the QCD laboratory with an adiabatically varying $\theta$-parameter; the amplitude of 
this variation exponentially (but still extremely slowly) grows in time.

 Eventually, as $\theta/N$ becomes larger than $\pi/N$, an observer in the cloud finds herself in a metastable reheated state. Initially the tunneling rate into the lower vacuum is much longer than the axion oscillation time (and, in fact,  longer than the age of the Universe), for instance, for $\theta/N$ close to $\pi/N$ it is given by the thin wall expression (\ref{p}). Oscillations continue in this regime until their amplitude 
 start approaching $\theta/N=\pi/2$, where mesons become tachyonic. Close to this saddle point the nucleation rate is determined by the thick wall result (\ref{G}) and becomes fast as $\theta$ approaches the saddle. When the nucleation rate becomes fast enough for a bubble of energetically favored vacuum to be created during one period of axion oscillations, the whole overheated part of the cloud with $\theta/N>\pi/N$ experiences the first order phase transition. 
 
Inside the transition region the latent heat of the overheated QCD vacuum is released  in the form of mesons. Note that the hidden sector with a dynamical axion has a single vacuum at $\theta=0$, so the net result of this process is that 
the axion cloud ``boils  up", and creates a cloud of strongly interacting mesons inside. Initially at the boundary of this cloud one has a ``$D$-wall" interpolating between two QCD vacua with $\theta=\pi$ (see, e.g., \cite{Gabadadze:2002ff} for a review). In the thick wall tunneling regime the energy carried by this wall is subleading with respect to the meson energy, and its presence is irrelevant for our purposes. The total energy released in mesons can be estimated as
\be
\label{energyrelease}
E_{mes}\sim m\Lambda^3r_g^3\sim {f_a^2\over M_{Pl}^2}M_{BH} \;,
\ee
where $m\Lambda^3$ is the latent heat of the phase transitions, approximately equal to the energy density in axions prior to the transition, and $M_{BH}$ is the black hole mass. Here we approximated the size of the cloud by $r_g$, which corresponds to setting $\alpha, l\sim 1$.

Mesons rapidly thermalize at a temperature 
\be
\label{temp}
T\sim (m\Lambda^3)^{1/4}\;,
\ee which is always higher than the meson mass $m_\pi\sim (m\Lambda)^{1/2}$, so one  ends up generating a gas of relativistic (or mildly non-relativistic)
strongly interacting particles.

Clearly, there are significant uncertainties in the estimates (\ref{energyrelease}), (\ref{temp}) but they clearly show that for high $f_a\sim M_{GUT}$ a significant ($10^{-5}\div10^{-3}$) fraction of a black hole mass can be converted into mesons by this mechanism.
Unlike axions, whose couplings are suppressed by the scale $f_a$ and extremely weak, hidden sector mesons may have appreciable interactions with the Standard Model fields, so this energy may be efficiently converted 
into photons. Before discussing more details  of this conversion let us comment on one possible caveat behind our reasoning.

The subtlety is that to trigger the instability the axion field needs to approach large enough values $\theta/N\sim\pi/2$. However, as discussed above, the main limiting factor for the growth of the axion cloud is the Bosenova collapse that happens at $\theta/N\sim 1$~\footnote{In fact, according to estimates of \cite{Arvanitaki:2010sy} it is likely to happen even earlier at small $\alpha/l$---at $\theta/N\sim 2\alpha/l$.}. Consequently, one may worry that the cloud collapse will always happen before the phase transition in the gauge sector, so that black holes can never trigger  hidden valley avalanches.

We believe this pessimistic conclusion is unlikely to be correct. First, even if the Bosenova collapse happens before the avalanche, it is very likely that the axion field significantly increases in the course of the collapse,
 so that the avalanche will still happen during the Bosenova event itself. To definitely confirm this expectation and to calculate the resulting energy release in pions a numerical simulation of the whole process is required, which is beyond the scope of our paper. So, as a proof of principle, let us describe an example of a situation where the avalanche definitely happens before the Bosenova collapse.
 
 To achieve this let us add two more parameters to our setup. First, we assume that the axion is coupled to the hidden valley topological density with an integer coefficient $N_a>1$,
 \[
 S_\theta={N_a\over 32\pi^2f_a}\int d^4x\;\phi_a\epsilon^{\mu\nu\lambda\rho}\Tr \,G_{\mu\nu}G_{\lambda\rho} \, .
 \]
Second, let us assume that the axion potential is dominated by a string contribution rather than by the hidden valley.
Then the axion potential takes the following form, 
\be
\label{existenceproof}
V=\Lambda_{st}^4\cos(\theta-\theta_0)-\Lambda^3m\cos\l N_a\theta/N\r\;,
\ee 
with $\Lambda_{st}^4\gtrsim \Lambda^3m$. Then the avalanche transition happens at $N_a\theta/N\sim \pi/2$, while axion non-linearities become important at $(\theta-\theta_0)\sim 1$. Clearly, there are  regions in this extended set of parameter, such that the avalanche happens in the regime when the 
axion field is still  linear. Note that in this case the energy released in mesons during the avalanche is suppressed by the ratio $\l \Lambda^3m/\Lambda_{st}^4\r$ as compared to the estimate (\ref{energyrelease}), so this ratio should not be too small.

Admittedly, this situation may appear more tuned compared to the one we were considering to start with. However, as we already stressed, we presented this specific example just as a proof of principle, and we expect that avalanches happen under more relaxed conditions as well.

\section{Electromagnetic Echoes of  Hidden Valley Avalanches}
\label{EM}
Let us turn now to discussing how the hidden valley mesons may be coupled to the Standard Model fields and what are the resulting electromagnetic signals in a different range of parameters.  There are several possibilities (``portals") of how a hidden valley  can communicate to the Standard Model fields (see, e.g., \cite{Batell:2009di}). For definiteness, we focus here on the vector portal, but the principal qualitative conclusions are model independent. 
In this case a subgroup of the flavor group of the
hidden QCD is gauged, so that some of the hidden sector mesons carry a charge under a new $U(1)$ boson $A'$, just like it happens for ordinary mesons. 

In principle, one may consider two different scenario of how $A'$ boson
interacts with the Standard Model fields.
The first possibility is that  some of the Standard Model fermions are charged under the $A'$, as considered, for instance, in \cite{Strassler:2006im}.  In this case the collider bounds constrain $A'$ to be relatively heavy, $m_{A'}/g'\gtrsim 10$~TeV. 

The more interesting possibility for our purposes  is that the only interaction between the Standard Model and  $A'$ is due to kinetic mixing with the hypercharge,
\be
\label{mix}
{\cal L}_{mix}=\epsilon F^Y_{\mu\nu}F'^{\mu\nu}\;,
\ee
where $F$'s are the hypercharge and $A'$ field strengths and $\epsilon$ is the mixing parameter. 
In this case $A'$ can be light or even massless if the mixing parameter $\epsilon$ is sufficiently small.
One theoretically motivated range for mixing is $\epsilon \sim 10^{-(3\div 2)}$.  Mixing parameters in this range would be generated at one loop 
by integrating out heavy fields charged under both $A'$ and the hypercharge \cite{Holdom:1986eq}. However,  mixing can naturally be much weaker as well. This would happen, for instance, if $A'$ is a part of a compact group, which is spontaneously broken at a scale much lower than the masses of all fields carrying both $A'$ and Standard Model charges. We will treat $\epsilon$ as a free phenomenological parameter. The summary of observational constraints on $\epsilon$ for different range of $A'$ and hidden pion masses can be found in, e.g., \cite{Jaeckel:2010ni}.

Let us see now how the constraints on the strength of interactions between a hidden sector and the Standard Model depend on the mass scale of a hidden gauge group. More precisely, the relevant parameter is
$\Lambda^3m$, which determines the overall scale of the hidden valley potential and, as a consequence,  the initial temperature of the meson cloud (\ref{temp}), the frequency scale of the resulting electromagnetic radiation, and also the value of the axion mass, $\mu_a^2\sim\Lambda^3m/f_a^2$.
Only relatively light sectors with $\Lambda^3m\lesssim (1\mbox{GeV})^4$ are relevant for our purposes, because axion masses for hidden valleys with larger energy scales are too heavy to trigger the superradiant instability
for astrophysical black holes.

Regardless of a specific communication portal there are two  different physical regimes, depending on the masses of hidden pions. Namely, pions lighter than a few MeV  may contribute as light species to the total energy budget during the Big Bang Nucleosynthesis (BBN). If the confinement scale $\Lambda$ itself is that low, the same problem arises due to hidden gluons and quarks.
To preserve the success of the BBN a hidden valley needs to be very weakly coupled to the Standard Model fields in this case, so that hidden mesons never reach thermal equilibrium after reheating 
in the early Universe. Typically,  there is also a strict  constraint from above on the reheating temperature after inflation, because at high temperatures thermalization through higher dimensional operators (for instance, related to heavy $A'$ exchanges) becomes more efficient.



For a mass scale of a hidden sector higher than $\sim 10$~MeV the constraints described above get significantly relaxed and interactions between the Standard Model and hidden sector fields may be much stronger.
In particular, it is possible for the hidden sector to be thermalized in the early Universe. In this case late decays of the hidden sector fields may cause distortions to the CMB black body spectrum or affect the BBN predictions if the life-time is too long. This issue does not arise if the life-time is shorter than $\sim 1$~s. 
This leads us to consider the scenario with a kinetically mixed light $A'$, otherwise the life-time gets too long for sub-GeV hidden valleys which we consider \cite{Strassler:2006im}. 
Then one can take $\epsilon$ as high as
$\sim 10^{-5}\div 10^{-3}$. A particularly interesting region of parameters
is when the $A'$ mass is also in the range  $\sim 10\div10^3$~MeV, which can also be probed in the near future
fixed-target experiments~\cite{Bjorken:2009mm}. This mass scale of the hidden sector 
corresponds to stellar mass black holes.

For lighter hidden sector mesons their mixing with the Standard Model should be very suppressed. The natural 
option to consider in this case is that the hidden photon is massless, so that hidden mesons acquire a tiny electric charge $\epsilon$ in the diagonal basis for photons. 
Observational bounds on the mixing parameter $\epsilon$ are very strict for light millicharged particles \cite{Davidson:2000hf,Jaeckel:2010ni}. For meson masses smaller than $\sim$few MeV the mixing parameter should be smaller than $\sim 10^{-9}$ from the cooling of SN1987A, for masses below $\sim$100 keV---smaller than $\sim 10^{-13}$ from the white dwarfs cooling, and smaller than $\sim 10^{-14}$ for masses below $\sim 10$~keV from red giants cooling.

Note, however, that  for light hidden sectors axions also gets lighter and as a result the size of black holes affected by superradiance increases. Consequently the total amount of emitted electromagnetic energy  can still be very significant. Another reason making this scenario potentially interesting is that the detectors
sensitivity generally increases when frequencies are getting lower.

The size of a cloud (in these estimates we do not distinguish
between the radii of a black hole and of a cloud)
 scales with the characteristic temperature $T\sim (\Lambda^3m)^{1/4}$ as
\be
\label{TRrelation}
R_c\sim\mu_a^{-1}
\leq
{f_a\over T^2}
\approx 20 R_{g\odot}  \l {f_a\over 2\cdot 10^{16}\mbox{ GeV}}\r
 \l {10\mbox{ MeV}\over T}\r^2
\ee
where $R_{g\odot}\approx 1.4$~km is the gravitational radius of the Sun. 
From this estimate we see that the most energetic photons originating from  pion decays and collisions may have frequencies ranging between $\sim 100$~MeV for stellar mass black holes and  $\sim 100$~eV for the largest supermassive black holes with masses of order $10^{10}M_\odot$.

Depending on the interaction strength and the size of the cloud, two evolution scenarios are possible. 
The distinction  is whether hidden sector particles are thermalized
with $e^+e^-$ and photons or not. The thermalization condition reads
\be
n\sigma R_c\sim\epsilon^2\alpha_{EM}TR_c 
\sim 7\cdot 10^{15}\epsilon^2 \l{T\over 100\mbox{MeV}}\r\l {R_g\over 2 \mbox{km}}\r\geq1\;.
 \label{ThermCond}
\ee
If this inequality is satisfied (which is possible for meson masses 
larger than $\sim$~100 keV) then the cloud behaves as a thermal gas including an electromagnetic component
with an initial temperature $T\sim (\Lambda^3 m)^{1/4}\sim (\mu f_a)^{1/2}$.
This gas
starts to expand and cool under its own pressure. 
So a likely outcome of the avalanche is a creation of a fireball consisting of photons and $e^+e^-$ pairs and carrying according to (\ref{energyrelease}) up to $\sim 10^{-4\div 3}$ of a black hole mass.

 Interestingly, this is similar to the starting point of one of the most successful model (the fireball model) for the $\gamma$-ray burst (see, e.g., \cite{Piran:2004ba,Meszaros:2006rc} for  reviews).  The required fireball energy in this model  is comparable to
 the avalanche energies in our setup, which is not surprising given that both energies are set by a rest mass of a stellar size object. 
Also the rate of avalanches is likely to be close to the rate of $\gamma$-ray bursts.
Indeed,
a significant fraction of $\gamma$-ray bursts is likely to originate from supernovae explosions leading to a birth of a rapidly spinning black hole \cite{MacFadyen:1998vz}. In turn, every spinning black hole is expected to experience several periods of superradiant activity per the age of the Universe in the presence of an appropriate axion.

There is an important difference between our setup and traditional ones, however. An important ingredient in the fireball model for $\gamma$-ray bursts is the baryonic ``contamination" of the electromagnetic plasma. This results in a significant delay of the photon signal, because the electromagnetic energy first gets converted into the kinetic energy of baryons and only later gets released into photons as a consequence of a shock formation. Given that superradiance may happen for a single black hole long after the initial supernovae explosion it is possible to have practically no baryon contamination in our case. The corresponding avalanches are likely to give rise to $\gamma$-ray burst-like events of very short duration and with a spectrum close to thermal. 

Another important property of conventional $\gamma$-ray burst  models is a strong collimation of electromagnetic energy into relativistic jets. The initial shape of a superradiant cloud is highly non-spherical, and it is
formed in a close vicinity of a rapidly rotating black hole, so it might give rise to a jet as well. Also, as we discussed, superradiance is likely to give rise to a series of repeating Bosenova collapses, so that these bursts may repeat themselves. 


Let us discuss now what happens if the thermalization condition (\ref{ThermCond}) is not satisfied (which 
must be the case for mesons lighter than $\sim$~100 keV). In this case
after mesons cool down they all decay into hidden photons, so the only electromagnetic signal  comes from the hot phase when ordinary photons are created in the hidden sector plasma by $\epsilon$-suppressed processes. As before, the life-time for the hot phase is set by the cloud size $R_c$~\footnote{If the hidden photon is massive with a life-time much longer than $R_c$,  there is also a delayed  electromagnetic signal from hidden photon decays.}. 

%
To estimate the resulting electromagnetic energy release in this case let us denote by $\Lambda_{\mu}$ the maximum energy scale of a hidden sector that can be reheated by an axion with a mass $\mu$,
\[
\Lambda_{\mu}\equiv (f_a\mu)^{1/2}\;.
\]
The actual temperature of mesons in the cloud can be lower than $\Lambda_{\mu}$ if the dominant contribution to the axion potential comes from some other source, as in   (\ref{existenceproof}). 
Then the electromagnetic energy release during the time interval $\sim R_c$ can be estimated as
\be
\label{Egamma}
E_\gamma\sim n^2\sigma R_c^4T\sim10^{-16} 
M_{\odot}
\l{\epsilon^2\alpha_{EM}\over 10^{-30}}\r
\l {M_{BH}\over M_\odot}\r^{3/2}
\l{100 f_a\over M_{Pl}}\r^{5/2}
\l{T\over\Lambda_{\mu}}\r^5\;, 
\ee
where we assumed that photons are produced predominantly as a result of bremsstrahlung, and estimated the corresponding thermal cross-section as $\sigma\sim\epsilon^2\alpha_{EM} T^{-2}$.

This estimate assumes that the total energy of produced photons is smaller than the total energy of the cloud 
\be
M_c\sim T^4R_c^3
\label{Mc}
\ee
which is indeed the case if the thermalization condition (\ref{ThermCond}) does not hold.

\begin{figure}[t!] 
 \begin{center}
 \includegraphics[width=5in]{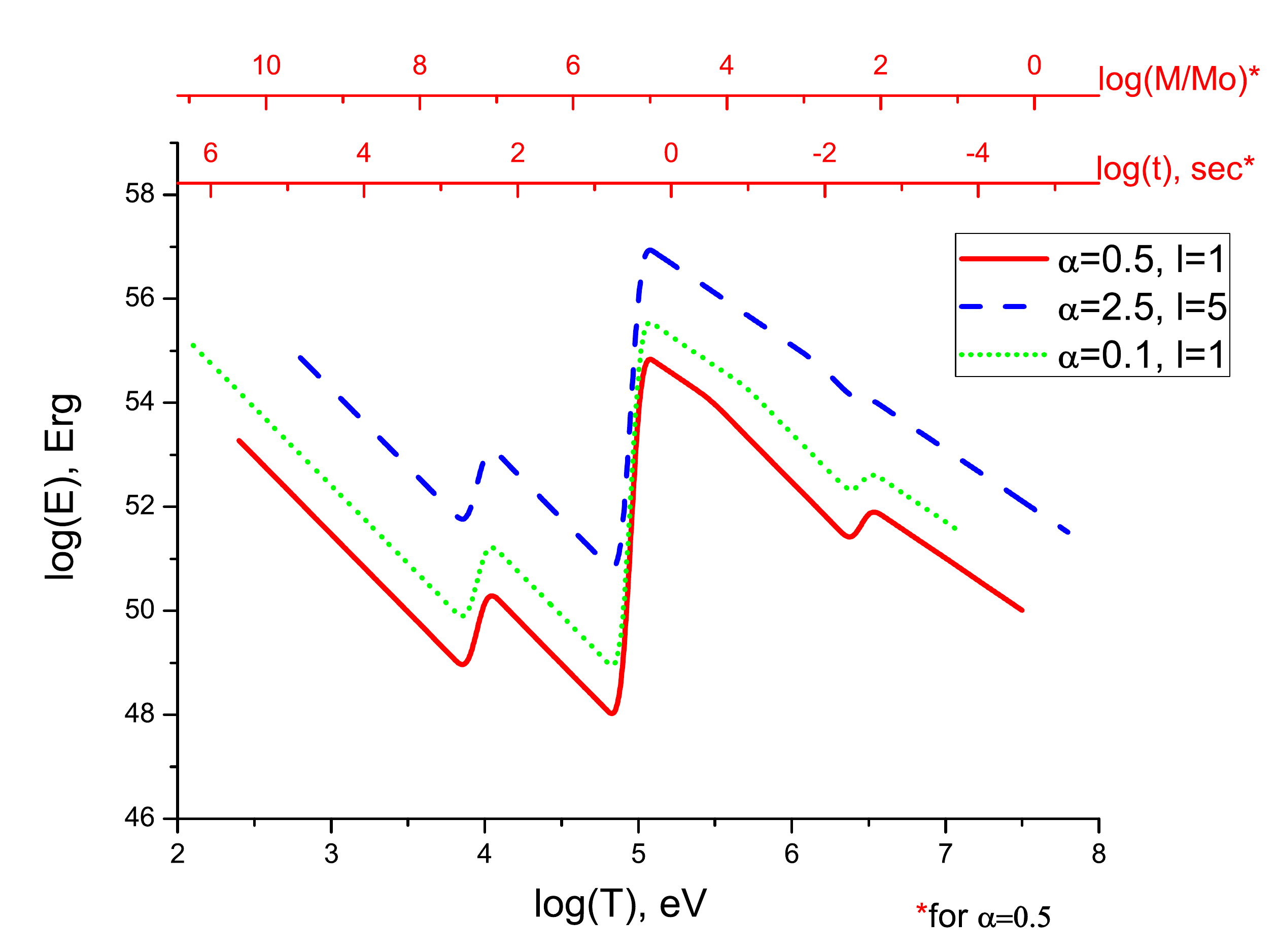}
\caption{A total energy release as a function of a characteristic temperature of 
a fireball for different values of parameters $\alpha$ and $l$. Two red axes on top represent
a corresponding black hole mass and time scale for  $\alpha=0.5$, $l=1$.
For the blue line a black hole mass is 5 times larger than the value on the red axis and a time
scale is 5 times longer. For the green line a black hole is 5 times lighter and a time scale is 
$\sim 5^{1/3}$ times longer. }
 \label{Eplot}
 \end{center}
\end{figure}

To illustrate the potentially observable effects of the hidden sector avalanches in Fig.~\ref{Eplot} we 
plot  an estimated total energy release as a function of the characteristic temperature of 
radiation. On the same plot we show the corresponding black hole masses and the characteristic
time scales. 

To produce this plot we assumed that the axion potential is dominated by the hidden sector, so that $T=\Lambda_\mu$ with $f_a=2\cdot 10^{16}$~GeV.
The total energy release is determined either by (\ref{Egamma}) or by the total mass of the axion cloud (\ref{Mc}) depending on whether thermalization condition holds or not. To check the latter we assumed that the mixing parameter $\epsilon$ takes the largest possible value compatible with the observational bounds at the corresponding meson mass scale.

The reason we have three different curves on the plot is the following. 
In our previous estimates we were assuming $\alpha\sim l\sim 1$. The three  curves are meant to illustrate  the uncertainties  in the estimates related to variations of  these two parameters.
The red solid curve on the plot corresponds to $l=1$ $\alpha=1/2$, {\it i.e.}, to  $\alpha\sim l\sim 1$ regime\footnote{Recall that the superradiance condition implies that $\alpha/l<1/2$.}. The dependence on $\alpha$ and $l$ comes mainly from two effects. First, the size of the cloud scales as $R_c \sim l^2/ \alpha $, affecting in turn the energy release through expressions  (\ref{Egamma}), (\ref{Mc}) and through the $R_c$-dependence of the thermalization condition. To illustrate this effect we presented the blue dashed curve, corresponding to $l=5$, $\alpha=2.5$.

Yet another effect enters when we vary $\alpha$ and $l$, such that the ratio $\alpha/l$ changes as well. Namely, as we already mentioned in section~\ref{avalanches}, the value of the $\theta$ parameter when the 
Bosenova collapse starts  was estimated in Ref.~\cite{Arvanitaki:2010sy} to be proportional to $\alpha/l$. Consequently, for small $\alpha/l$ the avalanche definitely does not happen before the Bosenova. Assuming that it still happens in the course of the Bosenova we can estimate the energy release in the following way. Let us take as a simplified model of an initial stage of Bosenova
that the axion cloud shrinks while preserving the total number of axions $N \sim \theta^2 R_c^3$.  If this is the case, $\theta$ inside the cloud grows and reaches  the value $\theta_c$
when the size of the cloud is 
$R'_c \sim {(\alpha/0.5)^{2/3}{R_c}}$.
Using this size instead of $R_c$ to estimate the energy release we obtain the dotted green curve on the plot, which corresponds to $l=1$, $\alpha=0.1$.
Clearly,  uncertainties  for this curve are even stronger than for our previous estimates.

\section{Discussion}
\label{conclusions}
In this paper we outlined a proposal how rotating astrophysical black hole may open a new window into hidden QCD-like gauge sectors coupled to light axions.
Thanks to axion superradiance rotating black holes may give rise to violent bursts of electromagnetic radiation with initial frequencies set by the confinement scale in the hidden sector.
Clearly, a number of details need to be filled in to extract concrete predictions for the characteristics of these sources,  allowing to distinguish them from astrophysical sources of a more conventional origin and to set limits on the existence of these phenomena at different frequencies.

First, our proposal relies on a qualitative scenario for the evolution of superradiance  instability developed in Ref.~\cite{Arvanitaki:2010sy}. Given that a black hole surrounded by
an axionic cloud is a very rich non-linear dynamical system it is highly desirable to confirm this scenario by detailed numerical simulations.

However, many features of our proposal can be studied without going into details of superradiance. Namely, from a purely phenomenological point of view what happens is that a black hole creates a fireball of photons and, if the frequencies are high enough, of electron/positron pairs. The initial temperature of the fireball is bounded from above by $\sim (f_a/R_g)^{1/2}$, where
$R_g$ is the black hole size and $f_a$ is the axion decay constant. This is a rather common situation in astrophysics, for instance the initial stage of the fireball model for $\gamma$-ray bursts is very similar. 

There is a number of differences in our case which affect the further evolution of the fireball and should help to distinguish this kind of sources. For instance, according to the results of Ref.~\cite{Arvanitaki:2010sy}, an episode of superradiant instability may happen for a single black hole in a medium with a very small baryon contamination.  In this case we expect the resulting burst to have spectrum close to thermal and have a shorter time scale than a typical $\gamma$-ray burst (here we refer to the case, when the frequency scale of the event corresponds to $\gamma$-rays). On the other hand, the electromagnetic 
fireball suggested here is immersed in a cloud of a dense liquid composed of strongly interacting hidden mesons, which may affect its evolution. 

When the episode of superradiance  happens to a black 
hole which was just created as a result of supernova explosion, the whole setup becomes really similar to conventional $\gamma$-burst models. Even in this case the details may be different, though.  For instance, 
superradiance is a relatively slow process. Furthermore, it can be shut off in a dense medium, so that it may be necessary for an environment to clean up a bit for the instability to start.
As a consequence we expect a significant time delay between a black hole collapse and the creation of a fireball.

Finally, supermassive black holes may give rise through superradiance to very powerful electromagnetic bursts at much lower (down to $\sim 100$~eV)  frequencies  than conventional $\gamma$-ray bursts.

To conclude, superradiance opens an intriguing opportunity to test new physics with astrophysical black holes.
Given that violent electromagnetic bursts are being observed in nature and the origin of some of them is unknown we  believe
a dedicated astrophysical study of the process proposed here is worthwhile.
\section*{Acknowledgments}
We thank Asimina Arvanitaki, Savas Dimopoulos, Gregory Gabadadze, Andrei Gruzinov, Albion Lawrence, and Andrew MacFadyen for useful discussions. 
\section*{Appendix: Extrema of the pion potenial are flavor diagonal}
Let us first show  that for non-degenerate quark masses all the extrema
correspond to diagonal $U$ matrices. 
%
By taking the derivative of (\ref{V(U)}) we arrive at the following extremality conditions,  
\begin{equation}\begin{split} 
	  Tr \left( i \left( \e^{-i {\theta}/{N}} M U  - U^{\dagger} \e^{i {\theta}/{N}}  M \right)	 H_a \right) = 0
  \label{dVtau2}
\end{split}\end{equation}
where 
 $H_a$  form an
orthonormal basis in the space of traceless Hermitian matrices. This condition implies that
\begin{equation}\begin{split} 
\e^{-i {\theta}/{N}} M U  - U^{\dagger} \e^{i {\theta}/{N}}  M
	   = i \tau {\mathbf 1} ,
  \label{dVtau3}
\end{split}\end{equation}
for some real number $\tau$. Taking the square of (\ref{dVtau3})
and substituting 
\[
MU M U = \e^{-2i\theta/N}M^2+i \tau\e^{-i\theta/N} MU ,
\]
which also follows from (\ref{dVtau3}), we get
\begin{equation}\begin{split} 
	   U_0 M^2 = M^2 U_0.
  \label{MUUM}
\end{split}\end{equation}
Clearly, if all elements of a diagonal matrix $M$ are different, a unitary
matrix also has to be diagonal to commute with it.

 Finally, if $k$ of the quark masses are degenerate, an extremum 
 may have a non-diagonal $k\times k$ block. However, this block 
  can  be diagonalized by a vector $SU(k)_V$ rotation of mesons, which is a symmetry of the chiral Lagrangian in this case.
 Consequently, we can always search for extrema in the diagonal form.

\end{document}